\documentclass[preprint,showpacs,aps,superscriptaddress,nofootinbib]{revtex4}\begin{document}

\title{Can  EPR correlations be driven by and effective wormhole?\footnote{Talk given at the  Eleventh Marcel Grossmann Meeting, Berlin, Germany, 23-29 July 2006.}}

\author{E. Sergio Santini}

\affiliation{Centro Brasileiro de Pesquisas F\'{\i}sicas,   
Coordena\c c\~ao de Cosmologia, Relatividade e Astrof\'{\i}sica ICRA-BR \\ 
Rua Dr. Xavier Sigaud 150, Urca 22290-180, Rio de Janeiro, RJ, Brasil
and\\
Comiss\~ao Nacional de Energia Nuclear \\ 
Rua General Severiano 90, Botafogo 22290-901, Rio de Janeiro, RJ, Brasil \\E-mail: santini@cbpf.br }


\begin{abstract}
We consider the two-particle wave function of an EPR system given by a two dimensional relativistic scalar field model. The Bohm-de Broglie interpretation is applied and the quantum potential  is viewed as modifying the Minkowski geometry. In such a way   singularities  appear in the metric, opening the possibility, following Holland,  of  interpreting the EPR correlations as originated by a wormhole effective geometry, through which physical signals can propagate.
\end{abstract}
\maketitle
A causal approach to the Einstein-Podolsky-Rosen (EPR) problem, i.e. a two-particle correlated system, is developed. We attack the problem from the point of view of quantum field theory  considering the two-particle function for a scalar field and  interpreting it according to the Bohm - de Broglie view. In this approach it is possible to interpret the quantum effects as modifying the geometry in such a way that the scalar particles see an effective geometry. For a two-dimensional static EPR model we are able to show that quantum effects introduces singularities in the metric, a key ingredient of a bridge construction or wormhole. Following a suggestion by Holland \cite{holland} this open the possibility of interpret the EPR correlations as driven by an effective wormhole\footnote{An extended version of this talk can be found in \cite{might} where a non-tachyonic  EPR model is studied }.

The two-particle wave function of a scalar field, $\psi_{2}(\bf{x_1},\bf{x_2},t)$ satisfies(see for example \cite{schweber}  \cite{long}):

\begin{equation}
\sum_{j=1}^{2}[(\partial^{\mu}\partial_{\mu})_{j}+m^2]\psi_{2}(\vec{\bf{x}}^{(2)},t) = 0
\end{equation}
where $\vec{\mathbf{x}}^{(n)}\equiv \{\bf{x}_{1},...\bf{x}_{n}\}.$ Explicitly we have

\begin{equation}\label{2p}
[(\partial^{\mu}\partial_{\mu})_{1}+m^2]\psi_{2}(\bf{x_1},\bf{x_2},t) + [(\partial^{\mu}\partial_{\mu})_{2}+m^2]\psi_{2}(\bf{x_1},\bf{x_2},t)= 0 .
\end{equation}

Substituting $\psi_2=R \exp(iS/\hbar)$ in Eq. (\ref{2p}) we obtain two equations, one of them for the real part and the other for the imaginary part. The first equation reads

\begin{equation}\label{2pr2}
\eta^{\mu_{1}\nu_{1}}\partial_{\mu_1}S\partial_{\nu_1}S
+\eta^{\mu_{2}\nu_{2}}\partial_{\mu_2}S\partial_{\nu_2}S = 2 {\cal M}^2
 \end{equation}
where

\begin{equation}\label{M2}
{\cal M}^2\equiv m^2\hbar^2 (1-\frac{Q}{2 m^2 \hbar^2})
\end{equation}
with $
Q \equiv Q_1+Q_2 $ being
$Q_1 = -\hbar^2\frac{ (\partial^{\mu}\partial_{\mu})_{1} R }{R}$
and $Q_2 = -\hbar^2\frac{ (\partial^{\mu}\partial_{\mu})_{2} R }{R} .\hspace{2cm}(4')$

The equation that comes from the imaginary part is

\begin{equation}\label{continu}
\eta^{\mu_{1}\nu_{1}}\partial_{\mu_1}(R^2\partial_{\nu_1}S)+\eta^{\mu_{2}\nu_{2}}\partial_{\mu_2}(R^2\partial_{\nu_2}S)=0
\end{equation}
which is a continuity equation.

Following De Broglie \cite{debroglie} we rewrite the Hamilton-Jacobi equation (\ref{2pr2}) as

\begin{equation}\label{2pre}
\frac{\eta^{\mu_{1}\nu_{1}}}{(1-\frac{Q}{2 m\hbar^2})}\partial_{\mu_1}S\partial_{\nu_1}S
+\frac{\eta^{\mu_{2}\nu_{2}}}{(1-\frac{Q}{2 m\hbar^2})}\partial_{\mu_2}S\partial_{\nu_2}S =
2 m^2\hbar^2 .
\end{equation}

Here $\eta^{\mu\nu}$ is the Minkowski metric and we can interpret the quantum effects as realizing a conformal transformation of the metric in such a way that the effective metric is
$g_{\mu\nu}= (1-\frac{Q}{2 m^2 \hbar^2})\eta_{\mu\nu}$.
Now, following an approach by Alves (see \cite{alves}), we will see that for the static  case it is possible to obtain a solution as an effective  metric which comes from Eqs. (\ref{2pr2}) and (\ref{continu}). 
For the static case these equations are:

\begin{equation}\label{2prtwo}
\eta^{11}\partial_{x_1}S\partial_{x_1}S + 
\eta^{11}\partial_{x_2}S\partial_{x_2}S = 2 m^2\hbar^2 (1-\frac{Q}{2 m\hbar^2})
\end{equation}

\begin{equation}\label{cont-two}
\partial_{x_1}(R^2\partial_{x_1}S)+\partial_{x_2}(R^2\partial_{x_2}S)=0
\end{equation}

We consider that our two-particle system satisfies the EPR condition $p_1=-p_2$ which in the BdB interpretation, using the Bohm guidance equation $ p=\partial_{x}S$, can be written as
$
\partial_{x_1}S=-\partial_{x_2}S .$
Using this condition in Eq. (\ref{cont-two}) we have
$
\partial_{x_1}(R^2\partial_{x_1}S)=\partial_{x_2}(R^2\partial_{x_1}S)$
and this equation has the solution
$R^2\frac{\partial S}{\partial x_1}= G(x_1+x_2)$
where $G$ is an arbitrary (well behaved) function of $x_1+x_2$. 
Substituting  in Eq.(\ref{2prtwo}) we have

\begin{equation}\label{qpot2}
 2 m^2\hbar^2 (1-\frac{Q}{2 m\hbar^2})= 2 (\frac{G}{R^2})^2
\end{equation}
and using the expression (4') for the quantum potential, the last equation reads
\begin{equation}\label{nonlinear}
8G^2 + (\partial_{x_1}(R^2))^2-2R^2\partial_{x_1}^{2}R^2 +(\partial_{x_2}(R^2))^2-2R^2\partial_{x_2}^{2}R^2 -8m^2 R^4 = 0 .
\end{equation}
A solution of this nonlinear equation is 
$
 R^4 = \frac{1}{2m^2}(C_{1} \sin(m(x_1+x_2)+C_2))
$
provided an adequated function $G(x_1+x_2)$ which can be obtained from (\ref{nonlinear}) by substituting the solution.

In order to interpret the effect of the quantum potential we can re-write Eq.  (\ref{2prtwo}) using (\ref{qpot2}) obtaining
$
m^2\frac{\eta^{11}}{(\frac{G}{R^2})^2}\partial_{x_1}S\partial_{x_1}S + 
m^2\frac{\eta^{11}}{(\frac{G}{R^2})^2}\partial_{x_2}S\partial_{x_2}S = 2m^2
$
that we write as

\begin{equation}
g^{11}\partial_{x_1}S\partial_{x_1}S + 
g^{11}\partial_{x_2}S\partial_{x_2}S = 2m^2
\end{equation}
and then  we see that the quantum potential was "`absorbed"' in the new metric $g_{11}$ which is:

\begin{equation}\label{m}
g_{11}=\frac{1}{g^{11}}=\frac{\eta_{11}}{m^2}(\frac{G}{R^2})^2=\frac{-\frac{C_{1}^2}{16 m^2}+\frac{3C_{1}^2}{16 m^2}\sin^2(m(x_1+x_2)+C_2)}{\frac{C_{1}}{2m^2} \sin(m(x_1+x_2)+C_2)} .
\end{equation}

We can see that this  metric is singular at the zeroes of the denominator in (\ref{m})and  this is characteristic of a two dimensional black hole solution (see \cite{mann} \cite{alves}). Then our two-particle system  "see" an effective  metric with singularities, a fundamental component of a  whormhole\cite{visser}. This open the possibility, following Holland \cite{holland}, of interpret the EPR correlations of the entangled particles as driven by an effective wormhole. Obviously a more realistic (i.e. four dimensional) and more  sophisticated model (i.e. including the spin of the particles) must be studied. 
\footnote{It is interesting to note that a wormhole coming from a  (Euclidean ) conformally flat metric with singularities was shown by  Hawking \cite{hawwormholes}. Consider the metric: 
\begin{equation}
ds^2=\Omega^2dx^2
\end{equation}
with
\begin{equation}
\Omega^2=1+\frac{b^2}{(x-x_{0})^2} .
\end{equation}
This looks like a metric with a singularity at $x_{0}$. However, the divergence of the conformal factor can be though as  the space opening out to another  asymptotically flat region connected with the first one by mean  a  wormhole of size $2b$.}  
\section*{Acknowledgements}
I would like to thank   Prof. Nelson Pinto-Neto, from ICRA/CBPF, Prof. Sebasti\~ao Alves Dias, from LAFEX/CBPF, Prof. Marcelo Alves, from IF/UFRJ,  and the 'Pequeno Seminario' of ICRA/CBPF for  useful discussions.
I would also like to thank Minist\'erio da Ci\^encia e Tecnologia/ CNEN and CBPF of Brazil for financial support.

\end{document}